# Extracting Contact Effects in Organic Field-Effect Transistors

Behrang H. Hamadani, and Douglas Natelson

***Abstract*— Contact resistances between organic semiconductors and metal electrodes have been shown to play a dominant role in electronic charge injection properties of organic field effect transistors. These effects are more prevalent in short channel length devices and therefore should not be ignored when examining intrinsic properties such as the mobility and its dependence on temperature or gate voltage. Here we outline a general procedure to extract contact current-voltage characteristics and the true channel mobility from the transport characteristics in bottom contact poly (3-hexylthiophene) field effect transistors, for both ohmic and nonlinear charge injection, over a broad range of temperatures and gate voltages. Distinguishing between the contact and channel contributions in bottom contact OFETs is an important step toward improved understanding and modeling of these devices.**

## I. Introduction

Recent developments in the area of plastic electronics have produced much excitement in the organic electronic community. Low-cost flexible displays, identification tags, logic for smart cards, etc have been envisioned as potential applications of organic electronic devices. Progress towards these objectives has intensified in the past few years due to dramatic improvements in organic semiconductor (OSC) synthesis, device mobility, film quality and contacts to these materials. However, a thorough understanding of the nature of transport in these devices and the physics of contacts is crucial to further development of optoelectronic organic devices.

Several studies have demonstrated that the contact resistance, $R_c$, plays an important role in transport characteristics of bottom-contact organic field effect transistors. In these studies, $R_c$ can easily dominate over the intrinsic channel resistance, $R_{ch}$, in short channel (few microns and below) devices [1-3]. The contact effects have to be taken into account when extracting the field effect mobility from the transport characteristics of the device; omitting these corrections gives an underestimate of the true channel mobility. Approaches used to differentiate between contact and channel resistances include analyses of single device characteristics [4,5], scanning potentiometry [6-8], and scaling of total device resistance with channel length in a series of devices [1-3, 9-11].

In a previous study [3], we reported measurements of $R_c$ and $R_{ch}$ as a function of temperature and gate voltage, $V_G$, in bottom contact poly (3-hexylthiophene) (P3HT) field effect transistors. These two components of the total source-drain resistance, $R_{on} \equiv \partial V_D / \partial I_D$, are extracted from the dependence of $R_{on}$ on the channel length, $L$. In that study, which used Au source/drain electrodes, we found that contact resistivity correlates inversely with mobility over four decades, over a broad range of temperatures and gate voltages. This is consistent with the predictions of a recent theory [12, 13] of organic semiconductor-metal contacts incorporating a thermionic emission model with diffusion-limited injection currents and accounting for the back flow of charge at the interface.

The method of extracting $R_c$ in Ref. 3 works well in devices where the OSC and the contacting metal electrodes form a linear, ohmic contact. In the case of nonlinear charge injection, one must consider the details of the relation between the injecting current and the contact voltage. In this paper, we outline a general procedure for extracting the contact current-voltage characteristics, $I_D - V_C$, in bottom contact OFETs. This approach, which we demonstrate to work regardless of the type of contact at the metal-polymer interface (i.e., ohmic vs. nonohmic), uses the scaling of the device current with channel length and the gradual channel approximation [5], to divide the total source-drain voltage, $V_D$ into a channel component and a voltage dropped at the contacts, $V_C$. We assume, as supported by scanning potentiometry [8], that $V_C$ is mainly dropped at the injecting

Manuscript submitted September 30, 2004. This work was supported in part by Robert A. Welch Foundation, the Research Corporation, and the David and Lucille Packard Foundation.

B. H. Hamadani is a graduate student with the Department of Physics and Astronomy, Rice University, Houston, TX 77005 USA

D. Natelson is an assistant professor in the Department of Physics and Astronomy and the Department of Electrical and Computer Engineering, Rice University, Houston, TX 77005 USA (correspondence should be addressed to email: natelson@rice.edu)

contact for metals with a significant Schottky energy barrier, $\Delta$. As expected, the $I_D - V_C$ characteristics extracted by this technique for a specific OSC-metal interface are unique at a given temperature and gate voltage, independent of $L$. We compare the (contact) field, temperature and gate voltage dependence of the injected current for charge injection in Au/P3HT devices, Cr/P3HT devices, Cu/P3HT devices, devices with interleaved Au and Cr source/drain electrodes. The temperature and gate voltage dependences of the channel mobility are also explored. Separating the contact and channel effects in bottom contact OFETs allows for improved understanding and modeling of the nature of charge injection in such devices.

## II. Experimental details

Devices are made in a bottom contact configuration [3] on a degenerately doped $p+$ silicon substrate to be used as a gate. The gate dielectric is 200 nm of thermal $SiO_2$. Source and drain electrodes are pattern using electron beam lithography in the form of an interdigitated set of electrodes with a systematic increase in the distance between each pair. The channel width, $W$, is kept fixed for all devices. Three different kinds of metallic electrodes (Au, Cr, Cu) were then deposited by electron beam evaporation followed by lift off. (25 nm of each, preceded by 2.5 nm of Ti adhesion layer; no Ti layer for Cr samples). Au electrodes were cleaned for one minute in a 1:1 solution of $NH_4OH$: $H_2O_2$ (30%), rinsed in de-ionized water, and exposed for about one min to oxygen plasma. The Cr samples were cleaned in the same manner followed by a last step dipping in a buffered HF solution for under 10 seconds. The HF is believed to etch the native oxide, exposing a fresh layer of dielectric. Cu electrodes on the other hand were only exposed to less than 25 seconds of $O_2$ plasma to clean the organic residue from the lift off. We found that Cu samples exposed to any cleaning procedure except for short $O_2$ plasma generally exhibited very poor transport properties.

The organic semiconductor is 98% regio-regular P3HT [14] a well studied material [15-17]. As received, RR-P3HT is dissolved in chloroform at a 0.02% weight concentration, passed through PTFA 0.02 micrometer filters, and solution cast onto the clean substrates, with the solvent allowed to evaporate in ambient conditions. The resulting films are tens of nanometers thick as determined by atomic force microscopy. The measurements are performed in vacuum (~ $10^{-6}$ Torr) in a variable-temperature probe station using a semiconductor parameter analyzer (HP4145B).

## III. Results and discussions

The devices operate as standard *p*-type FETs in accumulation mode [18,19]. With the source electrode grounded, the devices are measured in the shallow channel regime ($V_D < V_G$). Fig. 1a shows the transport characteristics ($I_D - V_D$) of an Au device with $L = 15\,\mu$m and $W = 200\,\mu$m at $T = 250$ K for a series of $V_G$s. The linearity of such plots is generally a signature of an ohmic contact between the P3HT and Au. Figure 1b shows the transfer characteristics ($I_D - V_G$) of the same device at different temperatures. This plot is an indicator of the on/off properties of the transistor. Note that we consider "off" to be the state of the device when no gate voltage is applied. The relatively poor on-off ratio is a result of slight doping of the P3HT upon exposure to air, which raises the $V_G = 0$ current.

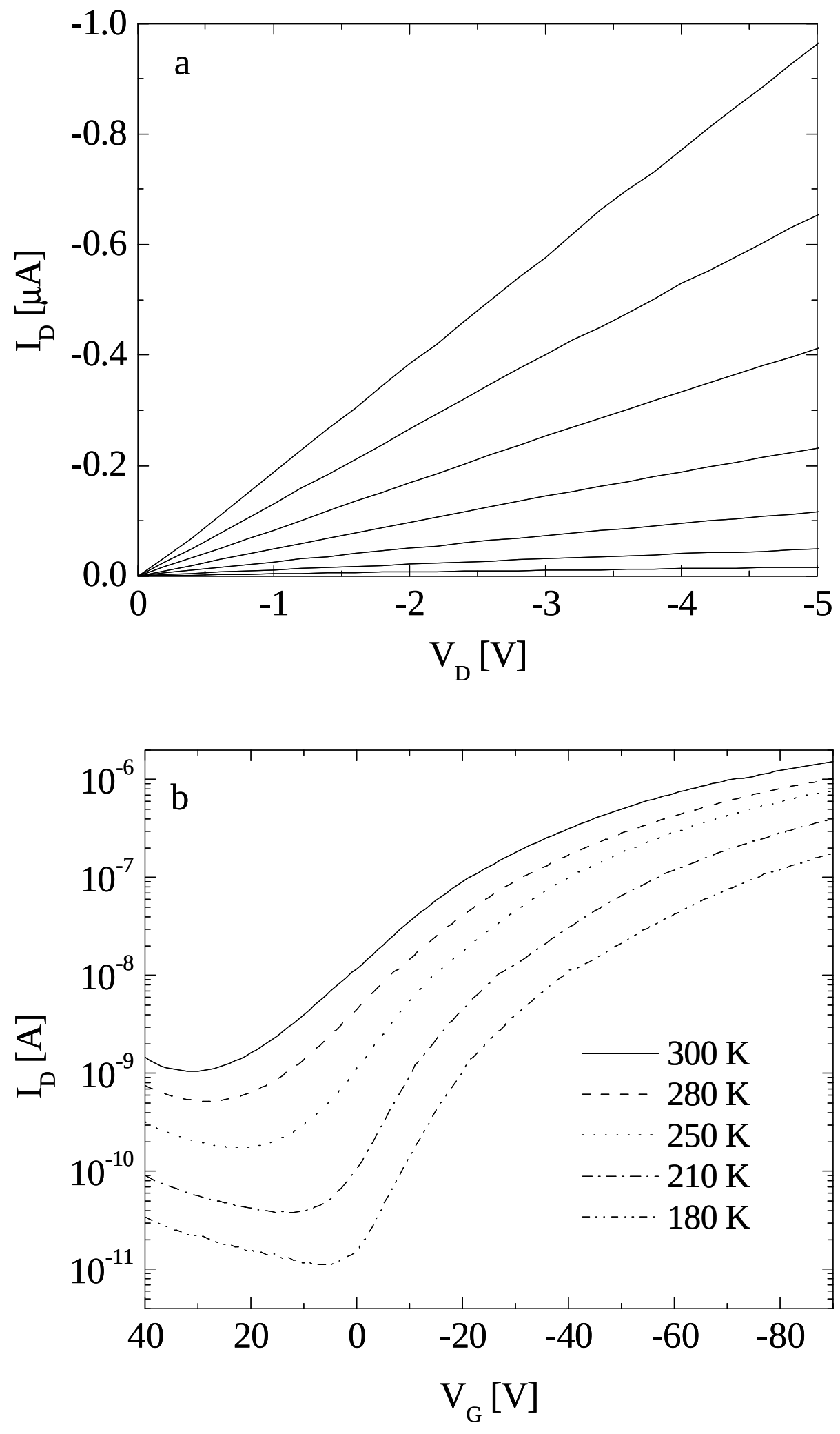

FIG. 1: (a) Transport characteristics of a P3HT OFET with Au source and drain electrodes, with $L = 15\,\mu$m, $W = 200\,\mu$m at 300 K. The $V_G$s from top to bottom are –70, -60, -50, -40, -30, -20, -10 V. (b) Transfer characteristics of the same device for several temperatures at $V_D = -3$ V.

The general approach to extract $R_c$ from plots such as those in Fig. 1a is to extract $R_{on}$ for each gate voltage and several channel lengths, while keeping $W$ constant. The slope in a plot of $R_{on}$ vs. $L$ describes the $R_{ch}$ per unit length of the channel and the intercept (the extrapolated resistance of a device of zero $L$) gives $R_c$. Fig. 2 shows data for a plot of $R_{on}$ vs. $L$ at different $V_G$s for an Au sample with $W = 200\,\mu\text{m}$ at $T = 300\,\text{K}$. From this plot, the contact and the channel resistance and therefore the true channel mobility $\mu$, can be extracted. For example, for the sample in Fig. 2, $R_c = 805\,\text{K}\Omega$ and $R_{ch} = 50\,\text{K}\Omega/\mu\text{m}$ at $V_G = -80\,\text{V}$ is extracted. From $R_{ch}$, a corresponding $\mu = 0.11\,\text{cm}^2/V.s$ is obtained. Repeating these measurements at different temperatures allows for the extraction of $\mu$ and $R_c$ as functions of temperature, therefore mapping out the general $R_c \propto \mu^{-1}$ relationship that exists in these devices over four decades of mobility [3].

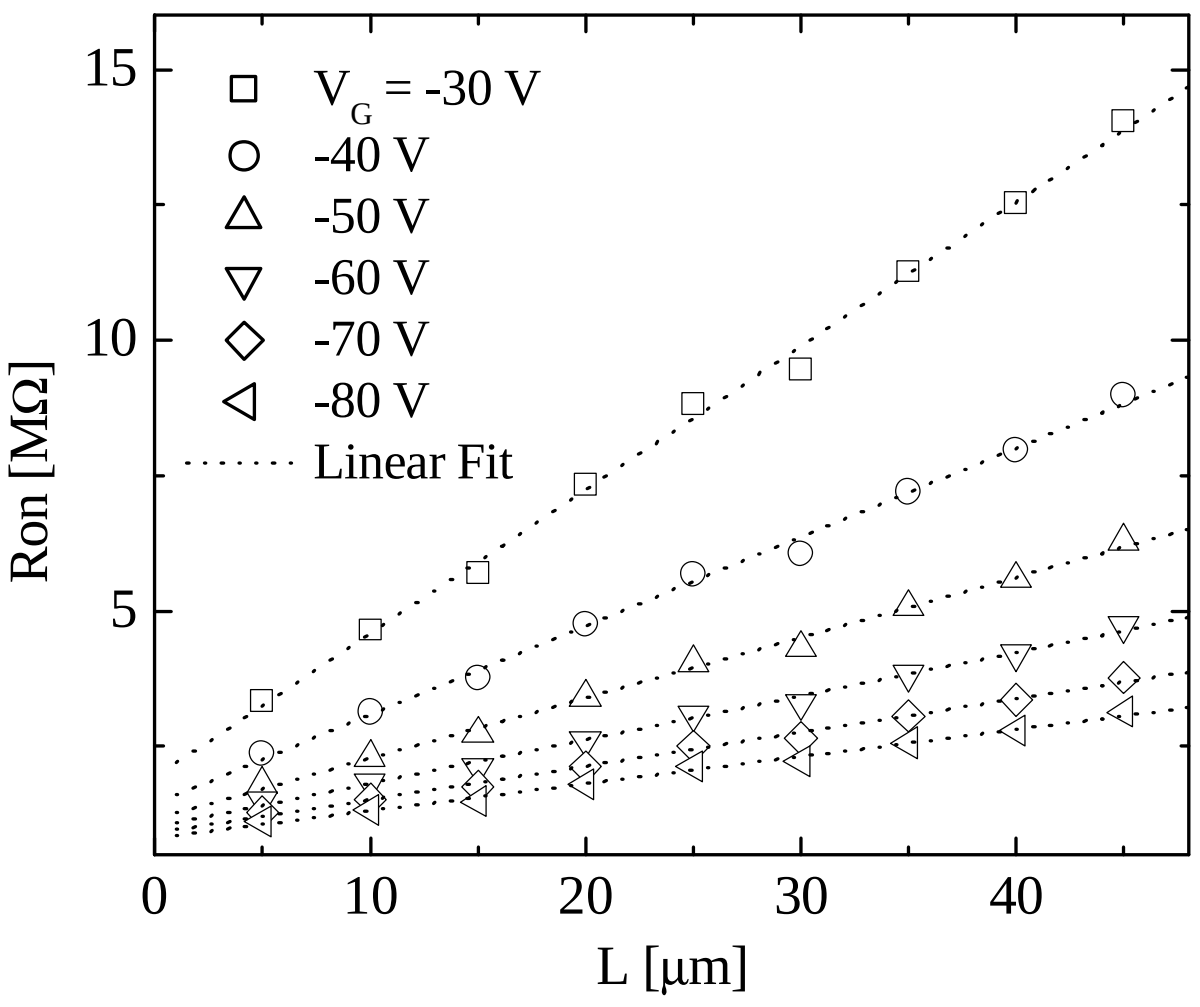

FIG. 2: $R_{on}$ as a function of L at 300 K for a series of Au-P3HT OFETs with channel width of 200 $\mu$m.

This technique of extracting $R_c$ from the transport plots works fairly well in samples with ohmic contacts to the OSC material such as the Au devices. In cases where a significant Schottky barrier is expected [8] to exist at the OSC-metal interface (such as in Cr or Cu/P3HT devices), the transport is nonlinear and one has to consider the details of the charge injection at the contact, i.e., $I_D - V_C$. The injection properties of the contact can be inferred [5] by splitting the channel into the main channel and a region near the contacts. A voltage of $V_C$ is assumed to be dropped across the contact portion, with the remaining $V_{ch} = V_D - V_C$ across the main channel. Using the charge control model [20], $I_D$ can be written as:

$$I_D = WC\mu[V_G - V_T - V(x)]\frac{dV}{dx}, \qquad (1)$$

where $V(x)$ is the potential in the channel at some position $x$, $V_T$ is the threshold voltage and $C$ is the capacitance per unit area of the gate dielectric. Integration of Eq. (1) from $x = 0$ to $L - d$ gives:

$$\frac{I_D}{WC\mu}(L-d) = (V_G - V_T)(V_D - V_C) - \frac{1}{2}(V_D^2 - V_C^2) \qquad (2)$$

where $V_C$ is dropped across $d$, a characteristics depletion length near the contacts. In this analysis, $L - d \approx L$ since $d << L$ in our samples (as it shall be discussed below, $d \sim 100\,\text{nm}$ while $L$ is tens of microns in all our samples). We also approximate that $V_C$ is dropped entirely at the injecting contact, i.e., the source electrode. Scanning potentiometry experiments [6-8] for this and other systems with a sizable $\Delta$ support this approximation. Finally, we set the threshold voltage to a constant 2 V in our analysis. Recent work [19] argues that $V_T$ is essentially ill-defined in such devices, since they operate without inversion. We note that varying $V_T$ by as much as a factor of five results in only a 13% change in the inferred mobility, without any apparent change in the shape of the extracted $I_D - V_C$ characteristics.

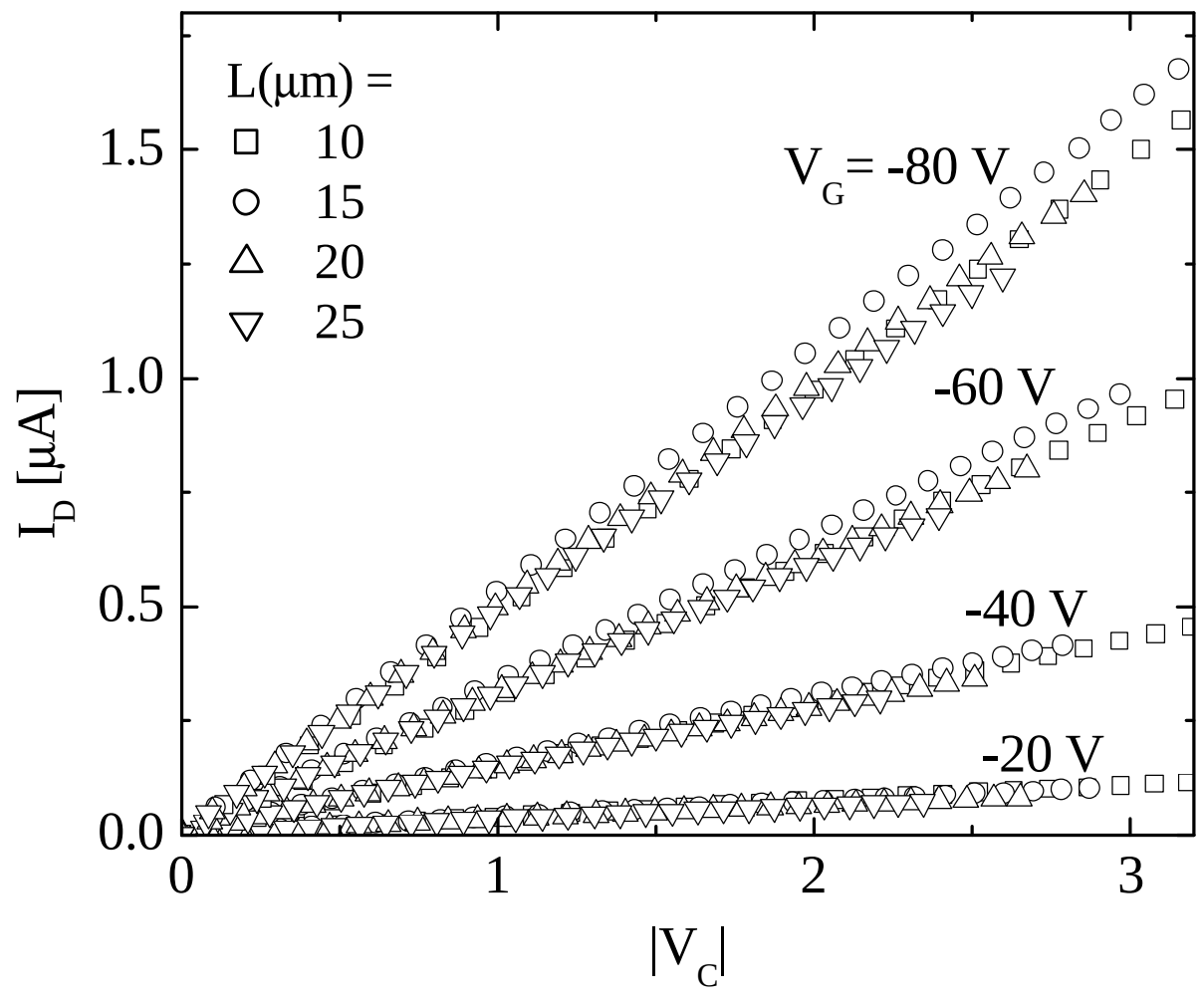

FIG. 3: Contact current voltage characteristics for the Au sample described in Fig. 2 at $T = 250\,\text{K}$. At any given $V_G$, the data for different channel lengths is collapsed onto one, using an appropriate value for $\mu$. Therefore, the $I_D - V_C$ data is unique for any $V_G$ and $T$.

For any given pair of $(V_D, I_D)$ data, a value of $V_C$ can be extracted from Eq. (2). This requires the knowledge of the correct value for $\mu$. To address this difficulty, we have used the length dependence of $I_D$ in an array of devices. At a given $T$ and $V_G$, a series of $I_D - V_D$ data is collected from devices with different channel lengths. The corresponding $I_D - V_C$ is calculated from Eq. (2) for all the different $L$. If the contact and channel transport properties in each device are identical, the correct value of $\mu$ would make all the different $I_D - V_C$ curves to *collapse onto one*; the injection characteristics of a particular OSC/metal interface should be unique and set by material properties and the (fixed) channel width and electrode geometry. Therefore the application of this technique results in simultaneous extraction of $\mu$ and $I_D - V_C$. We note that the source-drain field dependence of $\mu$ which is typically observed in organic semiconductors at high fields is not of a concern here, as the average source-drain field in our devices is low ($<10^3$ V/cm).

In order to test whether the method of Eq. (2) yields similar results to the technique of Fig. 2, we extracted $I_D - V_C$ characteristics of this Au sample. Fig. 3 shows the collapsed $I_D - V_C$ data for different gate voltages for the sample shown in Fig. 2 at $T = 250$ K. From this figure, it can be seen that the relationship between the current and the contact voltage is linear, consistent with presence of only a negligible energy barrier at the Au/P3HT interface. Also as expected, we find that the device parameters, such as $R_C$ and $\mu$ extracted from the collapse technique are *the same* as those obtained from $R_{on}$ vs. $L$ analysis. Fig. 4 shows the temperature dependence of $\mu$ extracted this way as a function of $T^{-1}$ for the same Au device. The *T*-dependence is well approximated as thermal activation consistent with simple hopping of carriers between localized states in the channel. The activation energies, $E_a$, are $V_G$ dependent, and they range from ~145 to 85 meV between the gate voltages of –20 to –80 V as it is seen in the inset of Fig. 4.

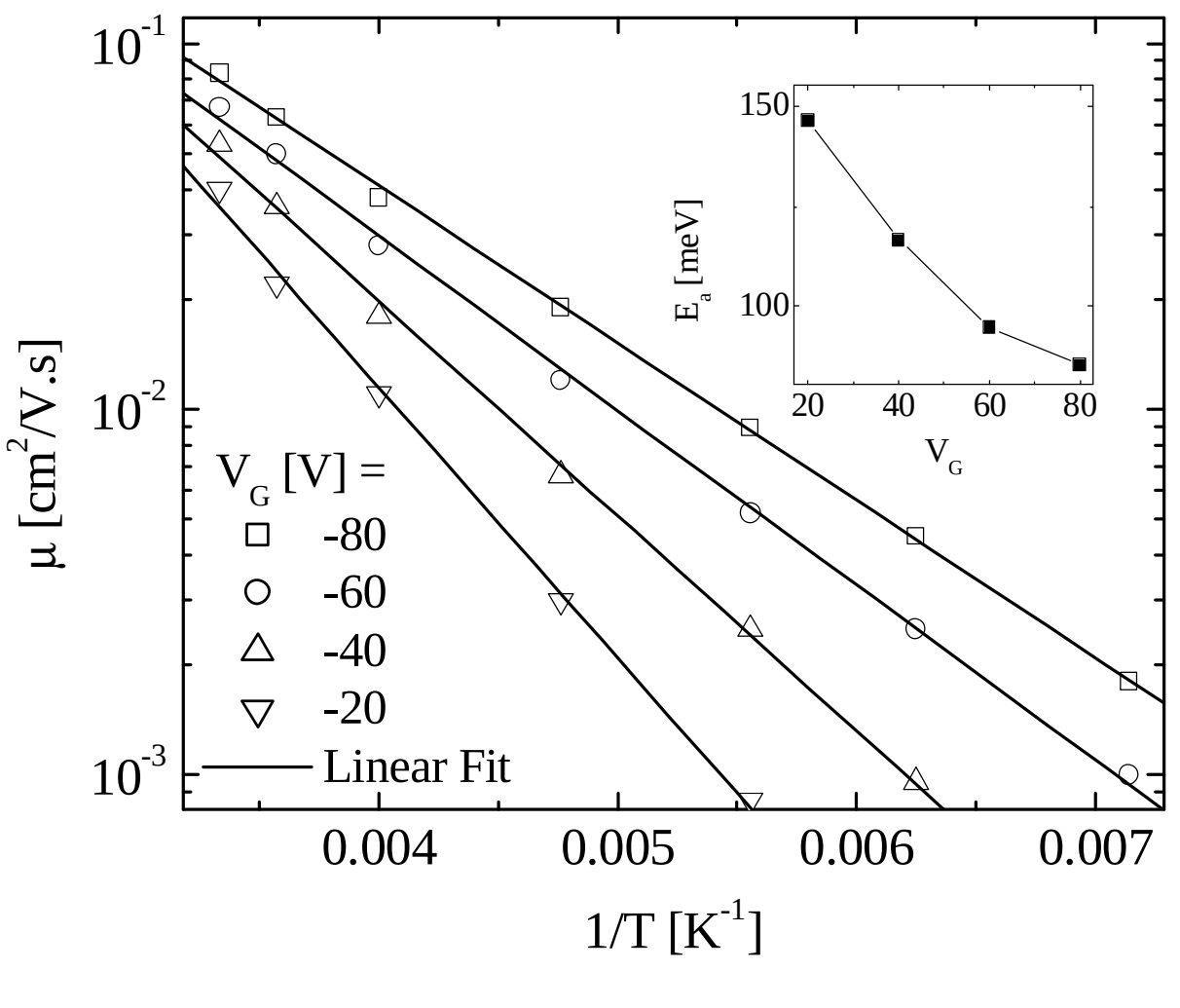

FIG. 4: Temperature dependence of the extracted channel mobility as a function of inverse temperature for a set of devices with Au source and drain electrodes at several gate voltages. Inset: Activation energies of the mobility as a function of the gate voltage

To confirm this method of extracting $\mu$ and $I_D - V_C$, we fabricated a series of devices (in a two-step lithography process) with alternating Au and Cr electrodes. The data is then taken twice for each device, once with the source electrode on Cr with the drain on Au and the second time vice versa. Fig. 5 shows a plot of the transport characteristics for injection into the channel from the Cr electrode in an Au/Cr sample with $L = 5\,\mu$m, $W = 200\,\mu$m at 240 K. The rise of current with voltage in the low $V_D$ regime ($V_D <$ 3 V) is close to quadratic. At higher drain voltages, the current becomes less nonlinear, at least in part due to saturation effects in the transistors. The severe nonlinearity observed in the injection data from Cr, in contrast to the linear injection from Au, is traditionally attributed [21,22] to the existence of a Schottky energy barrier, $\Delta$, between the Fermi level of the metal and the HOMO of the organic polymer.

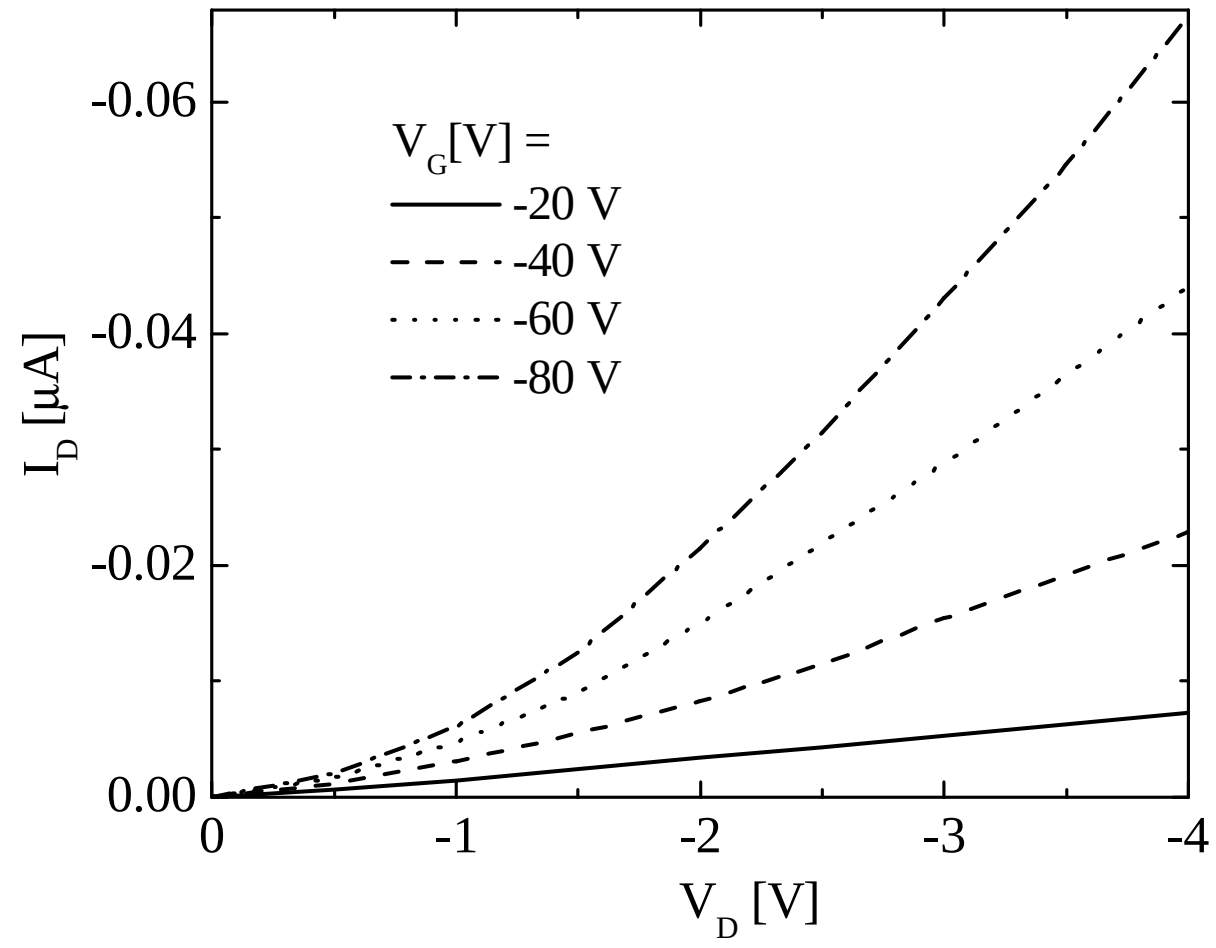

FIG. 5: Transport characteristics $I_D - V_D$ of an alternating Au/Cr OFET with source electrode on Cr. For this plot, $L = 5\,\mu$m, $W = 200\,\mu$m and $T = 240$ K.

Fig. 6 shows a plot of extracted $I_D - V_C$ for injection from Cr and Au at $T = 240$ K and $V_G = -80$ V. The plot indicates that charge injection from the Cr electrode is more severely contact limited at low $V_C$ while injection from Au allows for higher currents and lower contact resistance. We note that there is still a minute nonlinearity present in the data for injection from Au that is absent in the all-Au samples. We believe that this is consistent with a small contact effect at the drain, as was seen in the potentiometry profile of Cr/P3HT devices in Ref. [8]. This small voltage suppression at the drain does not appear to affect the transport at higher voltages as notably as the injection from the source does. The Au data in Fig. 6 have been shifted toward lower $|V_C|$ by 0.5 V to account for this drain voltage. The key observation for the non-identical source/drain samples is that the value of $\mu$ which collapses the different length-dependent data for injection from Cr onto one $I_D - V_C$ curve is *identical* to that inferred from the length-dependence of the channel resistance when injection is from Au in the same devices. This demonstrates that this process of extracting $I_D - V_C$ is well founded, and extracts the correct mobility.

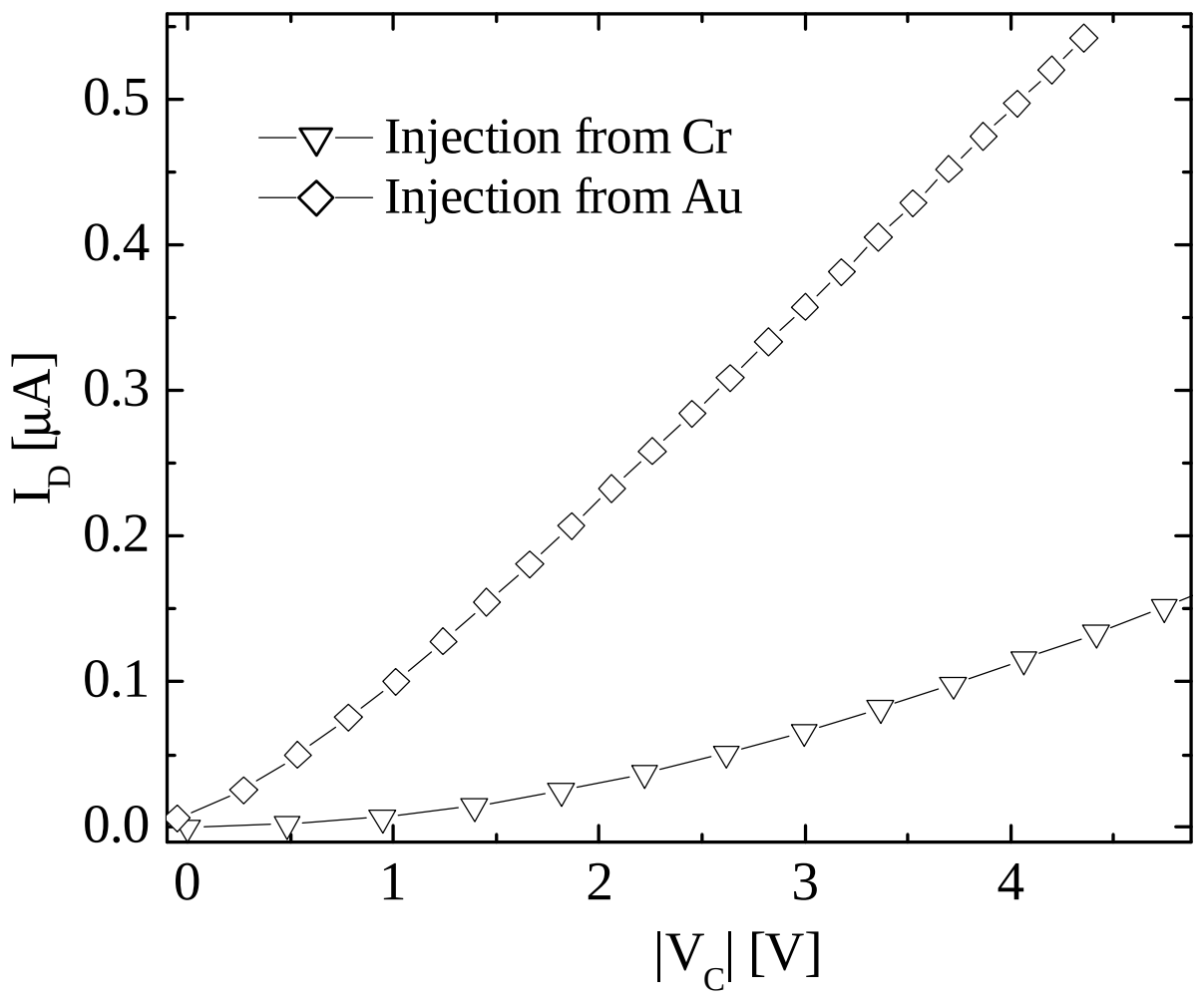

FIG. 6: Extracted $I_D - V_C$ for a series of devices of width $200\,\mu\mathrm{m}$ with alternating Cr and Au electrodes at 240 K and $V_G = -80\,\mathrm{V}$ . Upper curve shows injection of holes from Au, while the lower curve shows injection from Cr. Injection from Au is more linear and allows higher currents at lower voltages.

Fig. 7a shows a log-log plot of the extracted contact current-voltage characteristics of a set of Cr electrode devices at different temperatures and a fixed $V_G = -60\,\mathrm{V}$ . Fig. 7b shows the gate dependence of the injected current at a fixed *T*. These data are properties of the injecting contact metal-OSC interface, since the extraction procedure described above has removed the channel contribution. The inset shows the temperature dependence of the extracted channel mobility. As was observed in the Au devices, the *T*-dependence of the mobility in Cr samples is also described well by a single activation energy at any given gate voltage. Both the mobilities and the activation energies extracted from all-Cr devices are consistent (within typical device-to-device variation) with those seen in all-Au devices, as expected.

Fig. 8 shows a log-log plot of $I_D - V_C$ for a set of Cu electrode devices at $V_G = -60\,\mathrm{V}$ over a representative *T* range. The magnitude of the currents in our Cu samples has generally been lower than those in Cr devices. We have observed that the contact resistivity (the slope of $I_D - V_C$ in the low $V_C$ limit) increases from Au to Cr and from Cr to Cu. This observation is suggestive of an increase in the value of $\Delta$ at the P3HT/metal interface from Au to Cr to Cu. The activation energies of the mobility in all types of samples are similar, though the actual values of mobility at a given *T* and $V_G$ can differ for different samples. For example, the mobilities for Au and Cr devices were similar whereas $\mu$ in the alternating Cr/Au devices and also in Cu samples have generally been lower. We believe different factors, such as inferior surface cleanliness in the two-step lithography process (for the interleaved Cr and Au electrode samples), or the cleaning techniques (since we could not clean Cu samples the same way as Au or Cr were treated, as mentioned earlier) are responsible for differences in the values of the channel mobility in different devices. We also note that it is commonly observed [8, 24] the field effect mobility can be appreciably different in nominally identically prepared samples.

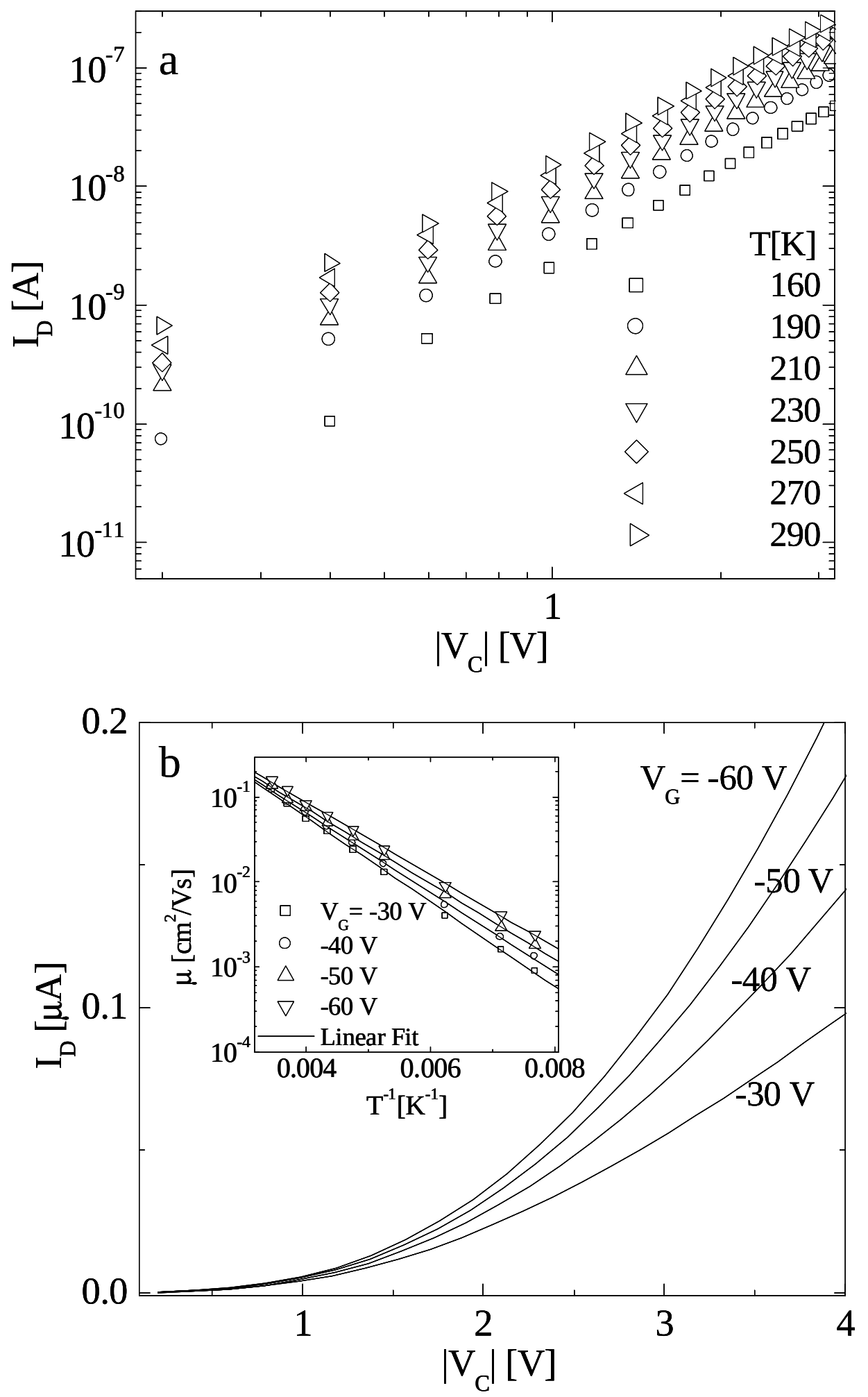

FIG. 7: Plots of extracted $I_D - V_C$ data from a set of Cr electrode devices at (a) fixed $V_G = -60\,\mathrm{V}$ and several temperatures and (b) $T = 210\,\mathrm{K}$ and several gate voltages. Inset in (b): Log-linear plot of $\mu$ vs. $T^{-1}$ for several gate voltages. The linear fit to the data indicates that the transport of carriers is dictated by simple hopping of carriers between localized states.

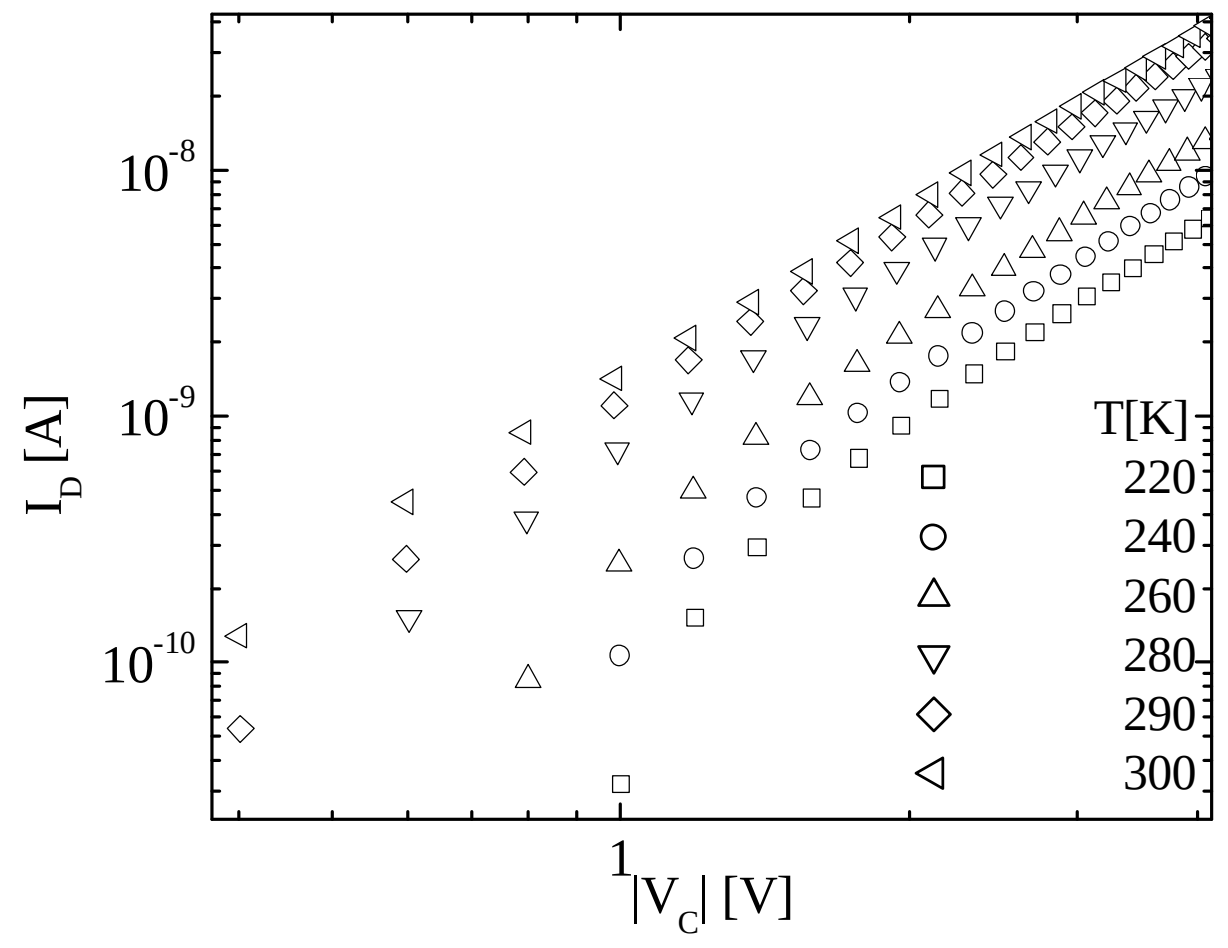

FIG. 8: $I_D - V_C$ for devices with Cu electrodes ($W = 400\,\mu$m) at $V_G = -60$ V over a representative temperature range.

One interesting feature of Figs 7a or 8 is the relatively weak temperature dependence of the injected current. At a fixed $|V_C|$ the temperature dependence of the injected current is reasonably described as thermally activated. The activation energies of the injected current in Cr at $|V_C| = 1$ V are *smaller* than the activation energies of the mobility by ~20 meV. These results are not consistent with the predictions of the traditionally accepted diffusion-limited thermionic emission models[12,13]. For Cr or Cu, $\Delta \sim 0.3$ eV [8] is expected to exist at P3HT/metal interface. Therefore, within the thermionic emission model the activation energy of the injected current in such samples is predicted to be $\sim (E_a + \Delta) \sim 0.4$ eV, which is clearly not the case for either Cr or Cu samples. These observations are in agreement with recent studies of charge injection in both bottom-contact P3HT field effect transistors [8] and hole injection from an Ag electrode into poly-dialkoxy-p-phenylene vinylene [23] where only a weak temperature dependence of contact resistance or injecting current was observed. These results imply that the diffusion-limited thermionic emission model is inadequate. A complete model of charge injection in these devices should account for both the temperature dependence of the charge injection as well as the correct field dependence of the current. We discuss this in more detail elsewhere [25].

Finally, an important issue to be addressed is the length scale and the origin of the characteristic depletion length, $d$ which appears in Eq. (2). The extent of this depletion region determines the magnitude of the electric field at the (injecting) contact. Scanning potentiometry experiments [8] and electrostatic modeling [26] have placed the extent of these regions at about 100 nm from the contact, depending on $V_G$. The origin of these regions in the vicinity of the contacts is not understood in detail, but they have generally been attributed to formation of regions of low carrier concentration (and of low mobility) near the contacts. Improved local probes (nm-resolution scanning potentiometry or cross-sectional scanning tunneling microscopy) would be extremely useful in better understanding these depletion regions.

## IV. Conclusions

Transport characteristics of a series of organic field effect transistors with P3HT as the active polymer layer and Au, Cr and Cu as the source/drain electrodes were examined over a broad temperature range. A general approach, incorporating the gradual channel approximation and scaling with channel length of the conductance, was used to extract the contact current-voltage properties of the different devices, even in situations where the contacts are not ohmic. This procedure was checked for consistency using devices with electrodes of alternating metal composition. This is a powerful technique for examining the injection properties of both ohmic and non-ohmic devices, as well as extraction of the true channel mobility in organic field effect devices.

**Behrang H. Hamadani** was born in Norman, OK in 1979. He has a BS in physics from the University of Texas at Dallas, and a MS in condensed matter physics from Rice University.

He is currently pursuing his PhD in the field of organic electronic devices in Prof. Natelson's experimental condensed matter physics group at Rice University in Houston, TX. He has published several papers and given talks on organic semiconductor devices.

Mr. Hamadani is a member of the American Physical Society and its Division of Polymer Science. He has been a recipient of the Ford Motor Company / Golden Key Honor Society scholarship award, and is being inducted into Sigma Pi Sigma, the physics honor society.

**Douglas Natelson** was born in Pittsburgh, PA in 1970. He has a BSE in mechanical and aerospace engineering from Princeton University (1993) and a PhD in condensed matter physics from Stanford University (1998).

After two years as a Postdoctoral Member of Technical Staff at Bell Laboratories, he joined the faculty as an Assistant Professor in the Department of Physics and Astronomy at Rice University in Houston, TX. In 2001 he was also made an Assistant Professor in the Department of Electrical and Computer Engineering. He has published recent papers on organic semiconductor devices, single-molecule transistors, and quantum coherence effects in metal nanostructures.

Prof. Natelson is a member of the American Physical Society and the Materials Research Society. He has been a recipient of the Research Corporation's Research Innovation Award, an Alfred P. Sloan Foundation research fellowship, a David and Lucille Packard Foundation fellowship, and a CAREER award from the National Science Foundation.